\documentclass[10pt,epsfig]{article}
\usepackage{amsfonts}

\usepackage{graphicx}
\usepackage{amsmath}

\input epsf.sty
\newtheorem{theorem}{Theorem}
\newtheorem{acknowledgement}[theorem]{Acknowledgement}

\input{tcilatex}
\makeatletter \@addtoreset{equation}{section}
\renewcommand{\theequation}{\thesection.\arabic{equation}}

\begin{document}

\title{\rightline{\mbox {\normalsize
{Lab/UFR-HEP0401/GNPHE/0404}}} \textbf{Algebraic Geometry Realization }\\
\textbf{of Quantum Hall Soliton}}
\author{R. Abounasr$^{1,3}$, M. Ait Ben Haddou$^{2,1,3}$,\ A.El Rhalami$^{1,3}$,
E.H.Saidi$^{1,3}$\thanks{%
h-saidi@fsr.ac.ma} \\
{\small 1 Lab/UFR-Physique des Hautes Energies, Facult\'{e} des Sciences de
Rabat, Morocco.}\\
{\small 2} {\small D\'{e}partement de Math\'{e}matiques \& Informatique,
Facult\'{e} des Sciences, Meknes, Morocco.}\\
{\small 3-Groupement National de Physique des Hautes Energies, GNPHE; Siege
focal, Rabat, Morocco.}}
\maketitle

\begin{abstract}
Using Iqbal-Netzike-Vafa dictionary giving the correspondence between the H$%
_{2}$ homology of del Pezzo surfaces and p-branes, we develop a new way to
approach system of brane bounds in M-theory on $\mathbb{S}^{1}$. We first
review the structure of ten dimensional quantum Hall soliton (QHS) from the
view of M-theory on $\mathbb{S}^{1}$. Then, we show how the D0 dissolution
in D2-brane is realized in M-theory language and derive the p-brane
constraint eqs used to define appropriately QHS. Finally, we build an
algebraic geometry realization of the QHS in type IIA superstring and show
how to get its type IIB dual. Others aspects are also discussed.

\end{abstract}

\thispagestyle{empty} \setcounter{page}{1}
\newpage
\tableofcontents
\newpage

\section{Introduction}

Few years ago, it has been conjectured in \cite{1}, see also \cite{2}, that
a specific assembly of a system of $KD0$, $D2$ and $N$ $D6$ branes and $N$
fundamental $F1$ strings, stretching between $D2$ and $D6$, has a low energy
dynamics similar to the fundamental state of fractional quantum Hall (FQH)
systems. There, the boundary states of the F1 strings ending on the D2 brane
are interpreted as the FQH particles moving in the D2 brane world volume.
The external strong magnetic field $B$ is represented by a large number of $%
D0$ branes dissolved in D2 and the dynamics of these particles is modeled by
a non commutative Chern-Simons (NCCS) $U(1)$ gauge field theory \cite{3}.
Soon after this proposal, several constructions were considered pushing
forward this analogy \cite{4}-\cite{12}. Susskind \textit{et al} idea was
also extended to quantum Hall solitons that are not of Laughlin type \cite
{13}, in particular the quantum Hall solitons modeling Haldane hierarchy and
multilayer systems as proposed in \cite{14}.

On other hand, it has been observed recently by Iqbal-Neitzke-Vafa (INV)
\cite{15} that there is remarkable correspondence between p-branes in
M-theory on torii and holomorphic curves in del Pezzo surfaces. A dictionary
characterizing this correspondence was given. The result of this work was
particularly focused on the study of a mysterious duality in the toroidal
compactification of M-theory, for other applications see also \cite{16,17}.
But here we will use differently the INV link between del Pezzos and
M-theory by developing a new method to approach brane systems. The
originality of our construction rests on the fact that INV correspondence
can be also used to study geometric aspects of brane physics using the power
of algebraic geometry and homology. Among our results, we quote the
derivation of new representations of p-brane systems using the H$_{2}$
homology of algebraic curves in del Pezzo surfaces.

To fix the idea on the way we will be doing things, we focus, in a first
step, on a special system of branes and show how new representations can be
built. The system we will be dealing with here concerns mainly the usual
quantum Hall soliton (QHS) we have introduced in the beginning of this
introduction. But in the discussion section, we will also draw the lines of
other constructions, in particular the way QHS is realized in type IIB
superstring on $\mathbb{S}^{1}$ as well as higher dimensional extensions.

The principal aim of this work is then to use results on ten dimensional
QHS, the INV correspondence as well as string theory and mathematical
results to develop new realizations of quantum Hall solitons using algebraic
geometry curves in del Pezzo surfaces. For simplicity, we will give here the
main lines of the method on Susskind \textit{et} \textit{al} QHS. A detailed
analysis and applications dealing with other types of brane systems will be
given in \cite{18}.

The presentation of this paper is as follows. In section 2, we describe
briefly the Quantum Hall Soliton in type IIA superstring. We first review
the structure of the ten dimensional QHS as formulated in literature and
then give its representation in the language of the eleven dimensional
M-theory on $\mathbb{S}^{1}$. This change from type IIA to M-theory allows
us to reach the two remarkable points: (i) give a geometric realization of
the standard idea of dissolution D0 branes in D2 and (ii) derive the
appropriate geometric constraint eqs that define QHS. In section 3, we
review the homology of del Pezzo surfaces as it is one of the basic tools in
construction and in section 4 we describe the INV correspondence. In section
5 we first identify the constraint eqs for QHS using the H$_{2}$ homology of
del Pezzos. Then we develop a realization of the Hall soliton using
intersecting classes of complex curves in del Pezzos. In section 6, we give
our conclusion and make a discussion.

\section{Quantum Hall Solitons}

One the remarkable observation of Susskind and collaborators in the
derivation of the Quantum Hall Soliton is that the usual $\left( 1+2\right) $
dimensional condensed matter Fractional Quantum Hall (FQH) phase has a
striking similarity with a specific p-brane configuration in type IIA
superstring theory. Following \cite{1,2}, see also \cite{19}, there is a one
to one correspondence between the 3d FQH systems of condensed matter physics
and the low energy dynamics of brane bounds involving D0, D2 and D6-branes
of the ten dimensional uncompactified type IIA superstring. There is also F1
strings stretching between D2 and D6 branes, F1 ends on D2 have an
interpretation in terms of FQH particles ( Hall electrons). Let us comment
briefly this configuration to which we shall refer here after as type IIA
stringy representation of quantum Hall soliton. Denoting the usual IIA
string (bosonic) coordinate field variables $X^{\mu }\left( \tau ,\sigma
\right) $ by the following equivalent and appropriate ones
\begin{equation}
\{t(\tau ),\qquad \varrho (\tau ,\sigma ),\qquad \vartheta (\tau ,\sigma
),\qquad \varphi (\tau ,\sigma ),\qquad \{y^{i}(\tau ,\sigma )\}_{4\leq
i\leq 9;}\},
\end{equation}
where $\tau $ and $\sigma $ are the usual string world sheet variables, the
above mentioned p-brane bound system, called also Quantum Hall Soliton
(QHS), is built as follows, see figure 1 for illustration:

\subsection{Brane Configuration}

If forgetting about edge excitations which may be modeled by NS5 branes, the
simplest structure of QHS is parameterized in terms of the above ten
dimensional string coordinates as follows:

\begin{figure}[tbh]
\begin{center}
\epsfxsize=10cm \epsffile{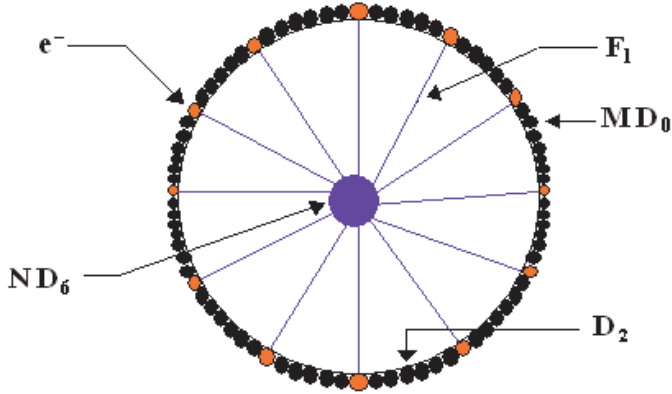}
\end{center}
\caption{\textit{This figure represents the type IIA stringy representation
of a fractional Quantum Hall Soliton. }}
\end{figure}
(\textbf{a}) One two space dimensional spherical D2 brane, it plays role of
the world volume in FQH systems of condensed matter physics and is
parameterized by the spherical coordinates,
\begin{equation}
\{t,\qquad \varrho =R;\qquad 0\leq \vartheta \leq \pi ;\qquad 0\leq \varphi
\leq 2\pi ;\qquad \mathbf{0_{6}}\}.
\end{equation}
At fixed time, this D2-brane is embedded in $R^{3}\sim R^{+}\times S^{2}$
and for large values of the radius, D2 may be thought locally of as $R^{1,2}$
which is also interpreted as the space time of the three dimensional Chern
Simons gauge theory.\newline
(\textbf{b}) $N$ coincident flat six dimensional space D6 branes located at
the origine of D2 and parameterized collectively by the $y^{i}$ internal
Euclidean coordinates as,
\begin{equation}
\{t,\qquad \mathbf{{0_{3}},\qquad }\left( y^{4}\mathbf{,}...,y^{9}\right)
\mathbf{\}.}
\end{equation}
It can be thought of as an external source of charge density $J^{0}\propto
N\delta ^{3}(x)$ at the origin $(x^{1},x^{2},x^{3})=(0,0,0)$ of the
spherical D2 brane.\newline
(\textbf{c}) $N$ fundamental strings F1 stretching between D2 and D6 branes
and parameterized by
\begin{equation}
\{t,\qquad 0\leq \varrho \leq R,\qquad \mathbf{{0_{2}},\qquad {0_{6}}\}}.
\end{equation}
The F1 string ends on the D2 brane are associated to the electrons of the
3-dimensional condensed matter FQH fluids.\newline
(\textbf{d}) $K$ D0-branes dissolved into the D2 brane; They define the flux
quanta associated to the external magnetic field $B$ of FQH systems. Recall
also that D0 and D6 are electric-magnetic dual.

\subsection{Methods}

Since the original work linking fractional quantum Hall fluids and NC
Chern-Simons gauge theory \cite{3}, several methods have developed to deal
with such kind of systems \cite{5,9}-\cite{20}. Matrix model approach \`{a}
la Polychronakos \cite{21} is one these methods which has been getting a
particular interest in literature. In this matrix model formulation, the FQH
particles in the Laughlin state are described by two $N\times N$ hermitian
matrices $X_{ij}^{1}(t)$ and $X_{ij}^{2}(t)$ ( in our coordinate choice $%
X_{ij}^{1}(t)\sim R\sin \vartheta _{ij}(t)\cos \varphi _{ij}(t)$ and $%
X^{2}(t)\sim R\sin \vartheta _{ij}(t)\sin \varphi _{ij}(t)$). For large
radius R, the two sphere can be locally approximated by a flat patch of the $%
\mathbb{R}^{2}$ plane and so one can neglect, in a leading approximation,
the curvature effect. In the infinite limit of $N$ and $M$ (strong external
magnetic field), the one dimensional matrix fields $X_{ij}^{1}(t)$ and $%
X^{2}(t)$ are mapped to the usual (2+1) fields, a behaviour which is nicely
given by Susskind map,
\begin{equation}
X^{i}(t,y)=y^{i}+\theta \varepsilon ^{ij}A_{j}(t,y)
\end{equation}
as discussed in \cite{3,10}. In this relation, one recognizes the $\left(
1+2\right) $ Chern Simons gauge field $A_{j}(t,y)$ and the non commutativity
parameter $\theta $ induced by the presence of external $B$.

In our present work, we will use a complete different approach to deal with
the QHS. This method is based on algebraic geometry of del Pezzo surfaces
and too particularly on their H$_{2}$ homology. In our way of doing, one may
naturally define all physical quantities one encounters in type IIA stringy
representation of QHS and condensed matter FQH fluids. For present
presentation however and in the purpose of illustration of the technique, we
will simplify the construction. We skip non necessary details and
essentially focus on the path towards the algebraic geometry realization of
QHS.

To proceed, let us say some words on our strategy towards the algebraic
geometry realization of QHS. This will be done in four principal steps: (i)
In step one,\ we reformulate the type IIA stringy representation of QHS as a
constrained system of p-branes. Here we show that the appropriate way to do
it is in fact from the view of M-theory on $\mathbb{S}^{1}$. In this case,
we give a geometric realization of the idea of dissolution of D0 branes in
D2 and show that QHS particles, namely electrons and flux quanta, can be
treated in a quite similar manner. This step permits us to identify the
appropriate geometric constraint eqs that define QHS. (ii) In step two, we
review the INV correspondence and describe how p-branes are represented in
algebraic geometry of del Pezzo surfaces. We take this opportunity to draw
the main lines of a method of representing homology classes in the del Pezzo
surface $\mathbb{B}_{1}$ by using F1 strings and D2 branes. This method uses
triangulation property of surfaces and is also motivated from formal
similarities with Feynman rules in quantum $\Phi ^{3}$ theory. (iii) In step
three, we reformulate the structure of the stringy QHS into the language of
homology of del Pezzo surfaces. We first give the translation of constraint
eqs in terms of H$_{2}$ homology of $\mathbb{B}_{1}$, then we study
necessary conditions for their solutions. (iv) In last step, we develop a
class of solutions of the homological constraint eqs giving an algebraic
geometry realization of QHS.

We begin by noting that p-branes involved in the above QHS may, roughly
speaking, be thought of as sets of points in p+1 dimensions. As far as brane
links are concerned, we clearly see that intersections between the QHS
branes may be naively defined as set intersections as follows:

\begin{equation}
D2\cap F1=1;\qquad D2\cap D6=0;\qquad D6\cap F1=1.  \label{cstr}
\end{equation}
For the case of N fundamental strings, the first\ equation of above
relations extends as $D2\cap \left( NF1\right) =N$ and so on. In ten
dimensional type IIA stringy representation, these relations are natural
identities that characterize the QHS and so they should be fulfilled in any
other representation of QHS including the algebraic geometry one we are
after. However to have a consistent description, we still need informations
about the K D0 branes of the QHS and which have no reference in eq(\ref{cstr}%
). This brings us to our first comment regarding this special property,
which to our knowledge have not been sufficiently explored in literature.
The idea of D0 dissolution in D2 is in fact strongly related with type IIA
representation of QHS requiring that the total space-time dimension of the
soliton should be equal to ten. However in eleven dimension M-theory on $%
\mathbb{S}^{1}$, we have an extra (compact) dimension which allows us to
engineer in a nice geometric way the D0 branes in perfect agreement with INV
correspondence. The key idea of our representation is summarized as follows:
The D0 branes ( flux quanta) dissolved in D2 are treated in M-theory on $%
\mathbb{S}^{1}$ on equal footing as the electrons in the sense that they
will be also viewed as ends of F1' strings, but this time, stretching
between D2 and K D0 branes, see figure1b for illustration.
\begin{figure}[tbh]
\begin{center}
\epsfxsize=10cm \epsffile{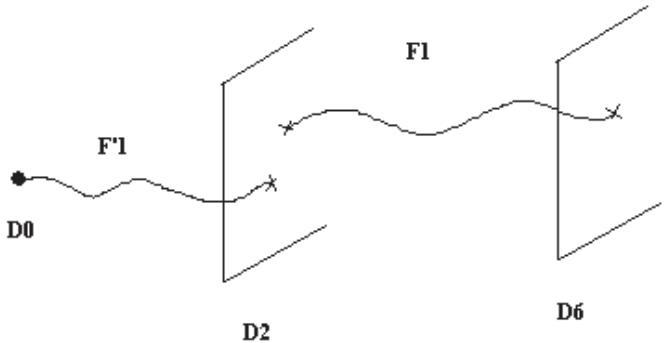}
\end{center}
\caption{\textit{This figure represents the type IIA stringy representation
of a fractional Quantum Hall Soliton. \ The ends of F'1 strings on D2 are
the D0 branes appearing in the Susskind el al QHS. They are in one to one
with the KD0s associated with the eleventh ( compact) dimension. }}
\end{figure}
From this representation, one clearly see that the total space time
dimension of the QHS is as in M-theory on $\mathbb{S}^{1}$. One also see
that D0 particles in QHS are associated with the compact direction $\mathbb{S%
}^{1}$ and moreover has much to do with the homological class of curve E$%
_{M} $ in del Pezzo surface $\mathbb{B}_{1}$ considered in \cite{15}. As
such we have, in addition to eq(\ref{cstr}), the following constraint eqs of
QHS formulated in the language of M-theory on circle $\mathbb{S}^{1}$,
\begin{eqnarray}
F^{\prime }1\cap D2 &=&1,\qquad F^{\prime }1\cap D6=0  \notag \\
D0\cap F^{\prime }1 &=&1,\qquad F^{\prime }1\cap F1=0  \label{cstr1} \\
D0\cap D2 &=&0,\qquad D0\cap D6=0,  \notag
\end{eqnarray}
where, leaving a part the brane dimension and their charge, there is a quite
similar analogy between the role of D0 and D6 branes. With this
reformulation of QHS in M-theory on $\mathbb{S}^{1}$ and to which we shall
continue to refer to it as type IIA stringy representation, we end step one
and are now in position to go ahead by following the drawn path. In the
second step, we describe briefly some useful tools on the H$_{2}$ homology
of del Pezzo surfaces and the INV correspondence between p-branes and
complex curves.

\section{Del Pezzo Surfaces}

In this section, we focus on two basic aspects. First we give some useful
tools on del Pezzo surfaces $\mathbb{B}_{k}$, $k=1,2,...,$ and too
particularly on the H$_{2}$ homology of their class of curves. Then we
consider the main lines of the toric representation of $\mathbb{B}_{1}$ as
this will be\ also relevant for later analysis.

\subsection{General on $\mathbb{B}_{k}$}

Del Pezzo surfaces $\mathbb{B}_{k}$ are complex dimension two compact
manifolds that are obtained by blowing up to eight points ($k\leq 8$) in
complex projective space $\mathbb{P}^{2}$ \cite{22,23}. These $\mathbb{B}_{k}
$\ complex surfaces are simply laced manifolds and their homology $%
H_{2}\left( \mathbb{B}_{k},Z\right) $\ is generated by the line class $H$ of
$\mathbb{P}^{2}$ and the exceptional curves $E_{i}$ generating the $k$ blow
ups of $\mathbb{P}^{2}$. The use of this line's homology turns out to be
very helpful in present study. It offers a poweful tool to study holomorphic
curves in del Pezzos and has the advantage of giving a quite complete
characterization of analytic curves without need to specify the explicit
form of complex algebraic geometry equations.

Recall that on a compact algebraic and projective variety X, a generic
divisor $\mathcal{D}=$ $\sum_{i}n_{i}\mathcal{D}_{i}$ is a finite formal
linear combination of complex co-dimension one analytic subvarieties $%
\mathcal{D}_{i}$. An instructive illustration of this construction is given
by the special case of a holomorphic function $F=F\left(
x_{1},x_{2},...\right) $ on X,
\begin{equation}
F=\dprod\limits_{j=1}^{s}F_{j}^{n_{j}}
\end{equation}
with $F_{j}=F_{j}\left( x_{1},x_{2},...\right) $ being the irreducible
components of F, they are holomorphic polynomials. Here the above $\mathcal{D%
}_{j}$s are the prime divisors associated with the zeros of $F_{j}$. The
divisor $\mathcal{D}$, which is called principal, reads as $\left( F\right)
=\sum_{j}n_{j}\left( F_{j}\right) $ with $n_{j}$ positive integers. The
support of the divisor is the variety V$\left( F\right) =\mathcal{D}_{1}\cup
\mathcal{D}_{2}\cup ...\cup \mathcal{D}_{s}$. Similar relations are also
valid for meromorphic functions with zeros and poles. Now, we turn to del
Pezzo surfaces and their homology.

In a given del Pezzo surface $\mathbb{B}_{k}$, each $E_{i}$ is associated
with a $\mathbb{P}^{1}$ holomorphic curve and the class system $\left\{
H,E_{i}\right\} $ satisfies the following pairing,
\begin{equation}
H^{2}=1;\qquad H.E_{i}=0;\qquad E_{i}.E_{j}=-\delta _{ij};\qquad i,j=1,...,k.
\label{bas}
\end{equation}
In terms of these basic classes of curves, one defines all the tools we need
for the present study. First, note that generic class of holomorphic curves
in $\mathbb{B}_{k}$ are given by linear combinations type,
\begin{equation}
C_{a}=n_{a}H-\sum_{i=1}^{k}m_{ai}E_{i},
\end{equation}
with $n_{a}$ and $m_{a}$ some integers. These classes of curves are
characterized by two basic parameters: (a) The self-intersection number $%
C_{a}^{2}$, which by help of eq(\ref{bas}) is given by,
\begin{equation}
C_{a}^{2}=n_{a}^{2}-\sum_{i=1}^{k}m_{ai}^{2},
\end{equation}
and (b) the degree $d_{C_{a}}$ which, as we shall see, is linked to the
space-time dimension of the p-branes. Since the $C_{a}^{2}$ and the degree
play a crucial role in the algebraic geometry realization of QHS we are
considering in this paper, it is interesting to note that among the above
classes of curves, there is a particular class of curves with a special
property. This concerns the canonical class $\Omega _{k}$ of the $\mathbb{B}%
_{k}$ surface which is given by minus the first Chern class $c_{1}\left(
\mathbb{B}_{k}\right) $ of the tangent bundle. It reads as,
\begin{equation}
\Omega _{k}=-\left( 3H-\sum_{i=1}^{k}E_{i}\right) ,
\end{equation}
and has a self intersection number $\Omega _{k}^{2}=9-k$ whose positivity
requires $k\leq 9$. Obviously $k=0$ corresponds just to the case where we
have no blow up; i.e the $\mathbb{P}^{2}$ complex surface. With the above
relation, we are now in position to define the degree $d_{C}$ of a given
curve class $C=nH-\sum_{i=1}^{k}m_{i}E_{i}$ in $\mathbb{B}_{k}$. It is the
intersection number between the class $C$ with the anticanonical class $%
\left( -\Omega _{k}\right) $,
\begin{equation}
d_{C}=-C.\Omega _{k}=3n-\sum_{i=1}^{k}m_{i}.
\end{equation}
Positivity of this integer puts a constraint equation on the allowed values
of the $n$ and \ $m_{i}$ integers which should be like $\sum_{i=1}^{k}m_{i}%
\leq 3n$. Note that there is a relation between the self intersection number
$C^{2}$ of the classes of holomorphic curves and their degrees $d_{C}$. This
relation, which is known as the adjunction formula, is given by
\begin{equation}
C^{2}=2g-2+d_{C},  \label{gen}
\end{equation}
it allows to define the genus $g$ of the curve class $C$ as $g=\left(
2+n\left( n-3\right) -\sum_{i=1}^{k}m_{i}\left( m_{i}-1\right) \right) /2$
where we have also used the expansion $C=nH-\sum_{i=1}^{k}m_{i}E_{i}$.
Fixing the genus $g$ to given positive number puts then a second constraint
equation on $n$ and $m_{i}$ integers. For the interesting example of
rational curves with $g=0$, we have then $C^{2}=d_{C}-2$ or equivalently
\begin{equation}
\sum_{i=1}^{k}m_{i}\left( m_{i}-1\right) =2+n\left( n-3\right) .
\end{equation}
For $k=1$, this relation reduces to $m\left( m-1\right) =2+n\left(
n-3\right) $, its leading solutions $n=1,m=0$ and $n=0,m=-1$ give just the
classes $H$ and $E$ respectively. Typical solutions for this constraint eq
are given by the generic class $C_{n,n-1}=nH-\left( n-1\right) E$ which is
more convenient to rewrite it as follows,
\begin{equation}
C_{n,n-1}=H+\left( n-1\right) \left( H-E\right)  \label{cur}
\end{equation}
In this case the degree of these rational curves in $\mathbb{B}_{1}$\ is
equal to $d_{C}=2n+1$, it deals then with p-branes in type IIA strings.
However along with the above solution, there is also configurations with
even degree. These solutions concerns NS branes given by the classes $%
C_{n,2-n}=C=nH-\left( 2-n\right) E$ with $n=1,2$ and wrapped p-branes in
type IIB representation. The second issue will be discussed later on.

\subsection{Toric representation of $\mathbb{B}_{1}$}

We need this toric representation to draw pictures for realizations of QHS
in terms of classes of curves in $\mathbb{B}_{1}$. To that purpose recall
first that toric representation is a tricky graphic representation that
concerns complex manifolds \cite{24,25}. The latters can be usually imagined
as given by a real base $\mathcal{B}$ with toric fibers on it. The simplest
examples toric manifolds are naturally the complex projective spaces $%
\mathbb{P}^{n}$ where the real dimension n bases $\mathcal{B}_{n}$ are given
by the usual n-simplex and fibers are n dimensional torii $\mathbb{T}^{n}$.
Therefore in toric representation $\mathbb{P}^{1}$ is an interval of a
straight line with a $\mathbb{S}^{1}$ circle on top and shrinking at the
boundaries. Similarly $\mathbb{P}^{2}$\ is a triangle with three vertices
capturing toric singularities. The blow up of one of these three toric
singularities of $\mathbb{P}^{2}$ is just $\mathbb{B}_{1}$ and is given by a
rectangle with four vertices but only two toric singularities. The
corresponding toric pictures of these three kinds of toric varieties are
shown on figures 2.
\begin{figure}[tbh]
\begin{center}
\epsfxsize=10cm \epsffile{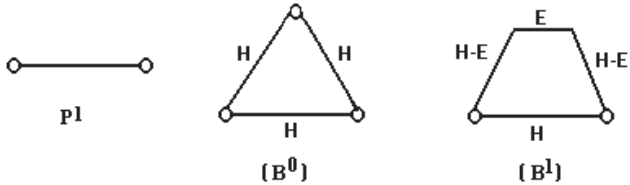}
\end{center}
\caption{\textit{Three} \textit{toric diagrams: (3a) Diagram of }$\mathbb{P}%
^{1}$\textit{\ with two }$\mathbb{S}^{1}$\textit{\ singularities on borders
of the intervalle. (3b) Toric graph of }$\mathbb{P}^{2}$\textit{\ with three
}$\mathbb{T}^{2}$\textit{\ toric singularities at the vertices and (3c)
toric diagram of }$\mathbb{B}_{1}$\textit{\ with two toric singularities and
a blown one.}}
\end{figure}
In the $H_{2}$ homology of del Pezzo surfaces where line classes in $\mathbb{%
B}_{1}$ are of two types, the H standard hyperline and the E exceptional
one, we have a nice description of these figures. Figures 2b and 2c are
respectively given by the following canonical lines of $\mathbb{P}^{2}$ and $%
\mathbb{B}_{1}$,
\begin{equation}
-3H;\qquad -\left( 3H-E\right) =-\left[ H+2\left( H-E\right) +E\right] .
\end{equation}
Naively, these canonical classes may be thought as representing the
boundaries of these complex surfaces, the triangle for $\mathbb{P}^{2}$\ and
rectangle for $\mathbb{B}_{1}$. Viewed in that way, these boundary lines are
genus one classes having degrees $d_{-3H}=9$ and $d_{-3H+E}=8$ respectively,
see eq(\ref{gen}). Moreover, the three edges of $\mathbb{P}^{2}$ (resp four
for the case of $\mathbb{B}_{1}$) correspond just to the number of
replication ( multiplicity) of the class H ( resp H and E for $\mathbb{B}_{1}
$) of the basis of the $H_{2}\left( \mathbb{B}_{k},\mathbb{Z}\right) $
homology. In other words the three ( resp four) edges for the toric graph of
$\mathbb{P}^{2}$ (resp $\mathbb{B}_{1}$) correspond to the splitting the
multiplicity $-3H$ as $-H-H-H$. The same is also valid for the three ( four)
vertices of the triangle ( rectangular), they correspond to the intersection
points of the classes of curves.

Along with these figures, one may also draw the pictures associated with the
rational curve classes of eq(\ref{cur}) inside the complex surfaces. Let us
give some illustrating examples which will be used later on.

\textbf{Graphs of the classes}\textit{\ }$H$ \textbf{and} $E$ \textbf{in} $%
\mathbb{B}_{1}$.

The hyperline class H has a self intersection one ($H^{2}=1$) and a degree $%
d_{H}=3$ giving the number of point on the boundary of $\mathbb{B}_{1}$. It
looks like a three point Feynman diagram with three external legs and a
three point vertex,
\begin{figure}[tbh]
\begin{center}
\epsfxsize=8cm \epsffile{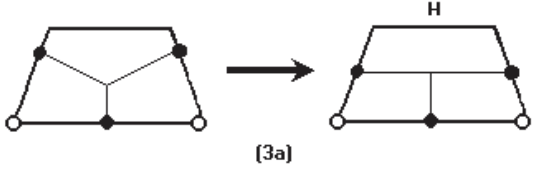}
\end{center}
\caption{\textit{The graph of the class H in }$\mathbb{B}_{1}$\textit{, it
looks like the Latin letter Y. But as the 3-vertex can be everywhere }$%
\mathbb{B}_{1}$\textit{, it can have different representation. In what
follows, we will use the block representation given on right. }}
\end{figure}
The unique self intersection point we have here belongs to the interior of
the $\mathbb{B}_{1}$ surface and may be interpreted as a signature of the H
class in $\mathbb{B}_{1}$, it is the 3-vertex of the triangulation of the
surface. For the exceptional curve E, one may be interested to do the same
as H. However this is not possible in H homological representation since E
has a negative self interaction ($E^{2}=-1$). This means \ that E cannot be
drawn inside of the rectangle. This is why we will avoid this behaviour by
changing the orthogonal basis $\left\{ H,E\right\} $ of the $H_{2}\left(
\mathbb{B}_{1},\mathbb{Z}\right) $ homology into the following equivalent
one
\begin{equation}
H,\qquad H-E  \label{nb}
\end{equation}
where the previous difficulty $E^{2}=-1$ is now solved as $\left( H-E\right)
^{2}=0$. The class of $H-E$ is a line in $\mathbb{B}_{1}$ with its two ends
on the boundary. Note that contrary to the old basis $\left\{ H,E\right\} $
which involves D0 and D2 branes, the new one implies instead F1 strings and
D2 branes. Note in passing that $\mathbb{B}_{1}$ may be also defined using
the following basis
\begin{equation}
l_{i}.l_{j}=1-\delta _{ij},\qquad i,j=1,2.
\end{equation}
In this basis, the canonical class reads as $-2l_{1}-2l_{2}$ and genus zero
curves of degree $2n+2$ are given by $C_{n,1}=nl_{1}+l_{2}$ or $%
C_{1,n}=l_{1}+nl_{2}$. They will be used later on when we consider type IIB
stringy representation of QHS.

\textbf{Graph of the class}\textit{\ }$H-E$ \textbf{in }$\mathbb{B}_{1}$

In the new basis eq(\ref{nb}) and thinking about the canonical class $\Omega
_{-3H+E}$ of $\mathbb{B}_{1}$\ as $-H-2\left( H-E\right) -E$, the class $H-E$
inside of $\mathbb{B}_{1}$is given by a line stretching between the basic H
and E classes of $\Omega _{-3H+E}$. This goes in the same manner as do the
two boundary ( external) lines $2\left( H-E\right) $ of the " line frontier"
class $\Omega _{-3H+E}$.
\begin{figure}[tbh]
\begin{center}
\epsfxsize=6cm \epsffile{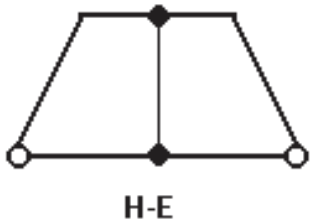}
\end{center}
\caption{\textit{The internal line crossing the surface of rectangle defines
the class }$H-E$\textit{. \ It may be viewed as a string stretching between
two boundary points on }$\mathbb{B}_{1}.$}
\end{figure}
The $H-E$ class has no\ self intersection ( no vertex).

\textbf{Graph of the class }$2H$ \textbf{in} $\mathbb{B}_{1}$

This is a genus zero class and has a degree equal to 6 and four self
interaction points, its picture is immediately obtained by summing the
graphs of two classes as $2H=H+H$. By superposition, we get in a first step
the figure 3c1,
\begin{figure}[tbh]
\begin{center}
\vspace{1cm} \epsfxsize=10cm \epsffile{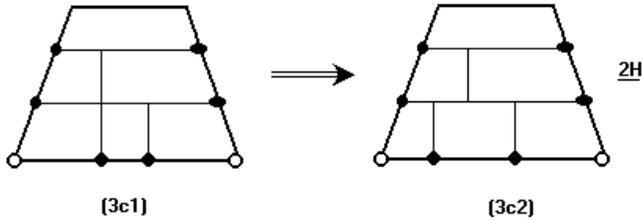}
\end{center}
\caption{\textit{Graph representing the class }$2H=H+H$\textit{. The
diagrams on left and right describe respectively the class 2H before and
after triangulation. In type IIA string language, this corresponds to NS5. }}
\end{figure}
which involves two kinds of internal vertices, a three vertex and a four
one. However splitting the four vertex into two 3-vertices using the the
following triangulation rule,
\begin{figure}[tbh]
\begin{center}
\epsfxsize=6cm \epsffile{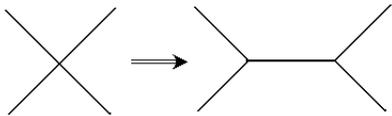}
\end{center}
\caption{\textit{These figures show the triangulation of four vertex A into
two three vertices A1 and A2.}}
\end{figure}
we get figure on right with the appropriate number of internal three
vertices. This property, which is general, is also valid for any sum of
class of curves.

\textbf{Graph of the class}\textit{\ }$2H-E$ \textbf{in} $\mathbb{B}_{1}$

Thinking about $2H-E$ as $H+\left( H-E\right) $ and following the same lines
as before, it is not difficult to show that the figure representing this
class of curve inside of $\mathbb{B}_{1}$ is given by.
\begin{figure}[tbh]
\begin{center}
\epsfxsize=8cm \epsffile{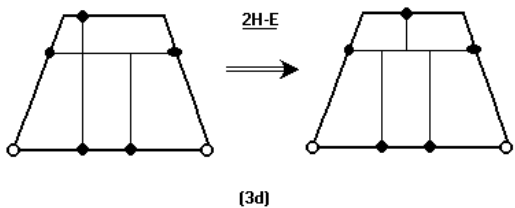}
\end{center}
\caption{\textit{Building block of the homology class 2H-E. It represents a
D4 brane in type IIA stringy description. }}
\end{figure}
Here also we recognize the five external legs and the three self
intersection points.

\textbf{Graph of the class}\textit{\ }$3H-2E$ \textbf{in} $\mathbb{B}_{1}$

Repeating the same process, we get for this class of curve, thought of as
the superposition of the following three basic curves $H+\left( H-E\right)
+\left( H-E\right) $, the graph of figure 3e
\begin{figure}[tbh]
\begin{center}
\epsfxsize=6cm \epsffile{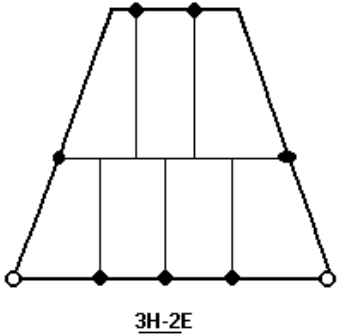}
\end{center}
\caption{\textit{Building block of the complex curve 3H-2E where one
recognizes the five self intersection points. It represents a D6 branes in
type IIA stringy description.}}
\end{figure}
Such procedure is general and applies for all classes of the H$_{2}$
homology of del Pezzos. Details will be given in \cite{18}

\section{Branes and holomorphic curves}

Following \cite{15}, there is a remarkable correspondence between del Pezzo
surfaces $\mathbb{B}_{k}$ and M-theory on $\mathbb{T}^{k}$. Generally
speaking an element $\omega $ of the\ real cohomology of del Pezzos
associates the basic classes $\left( H,E_{1},...,E_{k}\right) $ of the
surface $\mathbb{B}_{k}$ with the point $\left( l_{p},R_{1},...,R_{k}\right)
$ in the moduli space of M-theory on $\mathbb{T}^{k}$. In practice, this
means that $\omega $\ is a kind of generalized\footnote{$\omega $ has an
indefinite sign.} Kahler acting as,
\begin{equation}
\omega \left( H\right) =-3\ln l_{p};\qquad \omega \left( E_{i}\right) =-\ln
\left( 2\pi R_{i}\right) ,  \label{cc}
\end{equation}
where $l_{p}$ is the Planck scale and where $R_{i}$s are the torus radii. Eq(%
\ref{cc}) is in fact a special one, it happens that INV correspondence is
more general than given in eq(\ref{cc}). We have also the following
correspondences: (a) Global diffeomorphisms preserving the canonical class $%
\Omega _{k}$ of the del Pezzo surfaces corresponds precisely to U duality
group of M-theory on $\mathbb{T}^{k}$. (b) Rational curves (real two
spheres) $C$ with volume V$_{C}=\omega \left( C\right) $ and degree $%
d_{C}=\left( p+1\right) $ are in one to one with $\frac{1}{2}$\ BPS p-brane
states with tension $T_{p}=$2$\pi \exp V_{C}$. (c) Two classes of rational
curves $C_{1}$ and $C_{2}$ related as $C_{1}+C_{2}=-\Omega _{k}$ corresponds
just to the usual electric-magnetic duality linking Dp$_{1}$ and Dp$_{2}$
with $p_{1}+p_{2}=6$.

Therefore p-branes of ten dimensional IIA superstring can be realized as H$%
_{2}\left( \mathbb{B}_{k},\mathbb{Z}\right) $ homology classes of
holomorphic rational curves in $\mathbb{B}_{k}$. Of particular interest for
our present study is the realization of p-branes in terms of H$_{2}\left(
\mathbb{B}_{1},\mathbb{Z}\right) $ classes. More precisely, given a generic $%
\mathbb{B}_{1}$ rational curve $C_{n,m}=nH-mE$ with a positive degree $3n-m$
and integers $n$ and $m$ constrained as $m\left( m-1\right) =2+n\left(
n-3\right) $, we can work out all p-branes of type IIA superstring with
space dimension $p$ equal to $3n-m-1$. The result is reported on the
following table
\begin{equation}
\begin{tabular}{|l|l|l|l|l|l|l|l|}
\hline
Classes & C$_{0}$=E & C$_{1}$=H-E & C$_{2}$=H & C$_{4}$=2H-E & C$_{5}$=2H & C%
$_{6}$=3H-2E & C$_{8}$=4H-3E \\ \hline
Branes & D0 & F1 & D2 & D4 & NS5 & D6 & D8 \\ \hline
\end{tabular}
\ ,
\end{equation}
where now on the sub-index carried by the C$_{p}$s refers to the real space
dimension of the p-branes. From this correspondence, one sees that previous
figures we have drawn give indeed an algebraic geometry realization of
p-branes in terms of classes of holomorphic rational curves in $\mathbb{B}%
_{1}$. With these tools in mind, we are now ready to consider the main topic
of this paper.

\section{Realization of QHS}

To build a QHS representation using homology cycles of $\mathbb{B}_{1}$, we
start by recalling that form the type IIA string representation of QHS we
have the following first result,
\begin{equation}
\begin{tabular}{|l|l|l|l|l|l|}
\hline
p-Branes & $D0$ & $F^{\prime }1$ & $D2$ & $F1$ & $D6$ \\ \hline
Their realization in terms of Classes & $C_{0}$ & $C_{1}^{\prime }$ & $C_{2}$
& $C_{1}$ & $C_{6}$ \\ \hline
\end{tabular}
\ ,
\end{equation}
It gives the p-branes involved in QHS and their realization in terms of
classes of holomorphic curves in del Pezzo $\mathbb{B}_{1}$. Here $%
C_{1}^{\prime }$ refers to the class associated with fundamental strings
stretching between D0 and D2 and $C_{1}$ to those F1 strings stretching
between D2 and D6.

The next thing is to note that the problem of building algebraic geometry
realizations of QHS reduces then to the finding of explicit forms of these $%
C_{p}$ class of curves in terms of the $H$ and $H-E$ fundamental classes,
\begin{equation}
C_{p}=C_{p}\left( H,E\right) .  \label{sol}
\end{equation}
To do so, we first have to derive the appropriate constraint eqs that should
be obeyed by these $C_{p}$s, then solve them. We will see that a solution of
the form eq(\ref{sol}) that satisfy the QHS constraint eqs is not possible,
one needs much more ingredients which we describe at proper time.

\subsection{Constraint eqs and solution}

By identifying the notion of set intersection in real geometry with the
usual intersection of classes in H$_{2}$ homology of $\mathbb{B}_{1}$, the
constraint relations (\ref{cstr}-\ref{cstr1}) of type IIA string
representation of QHS translate in H$_{2}\left( \mathbb{B}_{1},\mathbb{Z}%
\right) $ homology language as follows:
\begin{eqnarray}
C_{0}.C_{2} &=&0;\qquad C_{0}.C_{6}=0;\qquad C_{2}.C_{6}=0,  \notag \\
C_{0}.C_{1}^{\prime } &=&1;\qquad C_{1}^{\prime }.C_{2}=1;\qquad
C_{1}^{\prime }.C_{6}=0,  \notag \\
C_{2}.C_{1} &=&1;\qquad C_{2}.C_{6}=0;\qquad C_{1}.C_{6}=1,  \label{wh} \\
C_{0}.C_{1} &=&0;\qquad C_{1}^{\prime }.C_{1}=0;\qquad C_{2}.C_{1}=1  \notag
\end{eqnarray}
At first sight, solving these constraint eqs for rational curves in del
Pezzo $\mathbb{B}_{1}$ seems a simple matter. However, this is no so
trivial. While the intersection of classes type $C_{2}.C_{1}=1$ or $%
C_{6}.C_{1}=1$ do cause no problem, the situation is not so obvious for the
constraint eqs $C_{2}.C_{6}=0$, $C_{0}.C_{2}=0$ and $C_{0}.C_{6}=0$. The
point is that there are no class of curves in $\mathbb{B}_{1}$ with such a
feature. This is easily seen by directly computing \ the corresponding
products. For instance the product between $C_{2}=H$ and $C_{6}=3H-2E$,
using eqs(\ref{bas}), gives,
\begin{equation}
C_{2}.C_{6}=3,
\end{equation}
and same thing for the other relations, which are not as required by the
structure of the QHS we are after. A way to over pass this difficulty is to
think about the three classes $C_{0}$, $C_{2}$ and $C_{6}$ as belonging to
three independent del Pezzo surfaces $\mathbb{B}_{1}^{\left( -1\right) },$ $%
\mathbb{B}_{1}^{\left( 0\right) }$ and $\mathbb{B}_{1}^{\left( 1\right) }$
as,
\begin{equation}
C_{0}^{\left( -1\right) }=E_{-1};\qquad C_{2}^{\left( 0\right)
}=H_{0};\qquad C_{6}^{\left( 1\right) }=3H_{1}-2E_{1},
\end{equation}
where in addition to eq(\ref{bas}), we also have $H_{0}.H_{\pm
1}=H_{0}.E_{\pm 1}=H_{\pm 1}.E_{0}=0$, see also figure 6. In this case, it
is not difficult to check that the intersection products $C_{0}^{\left(
-1\right) }.C_{2}^{\left( 0\right) }$, $C_{0}^{\left( -\right)
}.C_{6}^{\left( 1\right) }$\ and $C_{2}^{\left( 0\right) }.C_{6}^{\left(
1\right) }$\ are identically zero. The introduction of the $\mathbb{B}%
_{1}^{\left( -1\right) },$ $\mathbb{B}_{1}^{\left( 0\right) }$ and $\mathbb{B%
}_{1}^{\left( 1\right) }$ surfaces is the price one should pay for getting
solutions of QHS constraint eqs. As such one can think about these three
surfaces as three special sub-manifolds of the blown up of three different $%
\mathbb{P}^{2}$s\ embedded in $\mathbb{P}^{8}$. The two extra dimensions in $%
\mathbb{P}^{8}$ deal with the curves $C_{1}^{\prime }$ and $C_{1}$
associated with F'1 and F1 strings stretching between the two pairs $\mathbb{%
B}_{1}^{\left( -1\right) }-\mathbb{B}_{1}^{\left( 0\right) }$ and $\mathbb{B}%
_{1}^{\left( 0\right) }-\mathbb{B}_{1}^{\left( 1\right) }$ respectively.
Therefore a simple solution for the constraint eqs (\ref{wh}) read as
follows
\begin{eqnarray}
C_{0}^{\left( -1-1\right) } &=&E_{-1},  \notag \\
C_{1}^{\left( -10\right) } &=&H_{-1}-E_{0};\qquad C_{1}^{\left( 0-1\right)
}=H_{0}-E_{-1},  \notag \\
C_{2}^{\left( 00\right) } &=&H_{0},  \label{sol1} \\
C_{1}^{\left( 01\right) } &=&H_{0}-E_{1};\qquad C_{1}^{\left( 10\right)
}=H_{1}-E_{0},  \notag \\
C_{6}^{\left( 11\right) } &=&3H_{1}-2E_{1},  \notag
\end{eqnarray}
where the upper index $\left( ij\right) $ refers to the $\left( i,j\right) $
pair of the involved del Pezzos. The couple $\left( 00\right) $ (resp($%
\left( \pm 1\pm 1\right) $) means that we are dealing with classes of curves
in $\mathbb{B}_{1}^{\left( 0\right) }$ ( resp $\mathbb{B}_{1}^{\left( \pm
1\right) }$) and $\left( 0\pm 1\right) $ or $\left( \pm 10\right) $ with
rational curves stretching between $\mathbb{B}_{1}^{\left( 0\right) }$ and $%
\mathbb{B}_{1}^{\left( \pm 1\right) }$. Naturally, the full solution for
stretched F1 strings is given by the sum $C_{1}^{\left( 0\pm 1\right)
}+C_{1}^{\left( \pm 10\right) }$ which is equal to $\left( H_{0}-E_{\pm
1}\right) +\left( H_{\pm 1}-E_{0}\right) $.

In the above solution (\ref{sol1}) of the constraint eqs for QHS we have
considered one D6 brane and one D0 brane and same for the F1 string
stretching between D0-D2 and D2-D6. These p-branes are represented by $%
H_{2}\left( \mathbb{B}_{1}^{\left( 0,\pm 1\right) },\mathbb{Z}\right) $
classes describing rational holomorphic curves. In what follows, we derive
the general solution of eqs(\ref{wh}) involving ND6-branes and KD0-ones, N
and K are arbitrary positive integers.

\subsection{Quantum Hall Soliton}

The algebraic geometry realization of QHS built in terms of a system of
intersecting curves is as follows, see figure 6 for illustration

\begin{figure}[tbh]
\begin{center}
\epsfxsize=10cm \epsffile{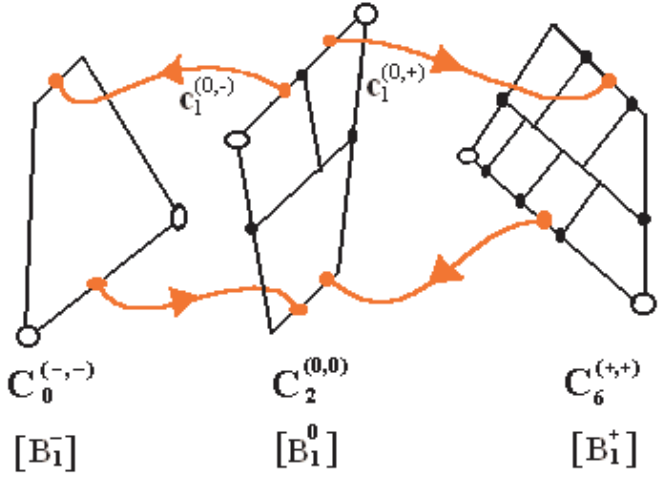}
\end{center}
\caption{\textit{In this figure, we give a configuration of the homological
representation of the QHS involving classes C}$_{0}$\textit{, C}$_{2}$%
\textit{\ and C}$_{6}$\textit{\ respectively associated with one D0, one D2
and one D 6 branes. We also represent the curves C}$_{1}$ \textit{%
illustrating strings stretching between the branes. }}
\end{figure}
(\textbf{1}) A simple rational curve class $C_{2}^{\left( 00\right) }=H_{0}$
belonging to a basic of del Pezzo copy denoted as $\mathbb{B}_{1}^{\left(
0\right) }$ and generated by $H_{0}$ and $E_{0}$ basic classes with same
properties as above. This class of curve is associated with the D2 brane in
type IIA stringy representation.\newline
(\textbf{2}) A class of curve with a multiplicity $N$ given by the class $%
C_{6}^{\prime }=N\left( 3H_{1}-2E_{1}\right) $. This is a non zero genus
class ($2g=N\left( 5N-7\right) +2$) that corresponds to the N coincident D6
branes in the IIA string representation of QHS. To see from where comes this
result, it is interesting to recall that in the case where the N D6 branes
are not coincident, the previous degenerate class split into N simple
classes of curves $C_{6}^{\left( ii\right) }$ given by
\begin{equation}
C_{6}^{\left( ii\right) }=3H_{i}-2E_{i},\qquad i=1,...,N.  \label{nd1}
\end{equation}
Each one of these N $C_{6}^{\left( ii\right) }$s belongs to one of the N del
Pezzo copies $\mathbb{B}_{1}^{\left( i\right) }$. The latters have basis $%
\left\{ H_{i},E_{i}\right\} $ with same properties as before but orthogonal
to the $\left\{ H_{j},E_{j}\right\} $ whenever $i\neq j$. This means that in
H$_{2}$ homology, the N D6 branes involve N copies of del Pezzo surfaces $%
\mathbb{B}_{1}^{\left( 1\right) },...,\mathbb{B}_{1}^{\left( N\right) }$ and
so requires a larger embedding projective space. To have an idea on the
dimension of this space, note that, in addition to F1 string loops $%
C_{1}^{\left( ii\right) }=\left( H_{i}-E_{i}\right) $ emanating and ending
on the same $C_{6}^{\left( ii\right) }$ copy, we have moreover other F1
strings stretching between the D6 branes. In H$_{2}$ homology, this
correspond to curves $C_{1}^{\left( ij\right) }$ and $C_{1}^{\left(
ji\right) }$ stretching between $\mathbb{B}_{1}^{\left( i\right) }$ and $%
\mathbb{B}_{1}^{\left( j\right) }$. The explicit expression of these classes
is given by
\begin{eqnarray}
C_{1}^{\left( ij\right) } &=&\left( H_{i}-E_{j}\right) ;\qquad C_{6}^{\left(
ii\right) }.C_{1}^{\left( ij\right) }=1;\qquad C_{1}^{\left( ij\right)
}.C_{6}^{\left( jj\right) }=1  \notag \\
C_{1}^{\left( ji\right) } &=&\left( H_{j}-E_{i}\right) ,;\qquad
C_{1}^{\left( ji\right) }.C_{6}^{\left( ii\right) }=1;\qquad C_{6}^{\left(
jj\right) }.C_{1}^{\left( ji\right) }=1.  \label{nd2}
\end{eqnarray}
From these relations, one clearly see that the $C_{1}^{\left( ij\right) }$
and $C_{1}^{\left( ji\right) }$ classes are stretching between the $%
C_{6}^{\left( ii\right) }$\ and $C_{6}^{\left( jj\right) }$. Since these $%
C_{1}^{\left( ij\right) }$ and $C_{1}^{\left( ji\right) }$ classes require
at least one complex dimension, the embedding projective space should be at
least $\mathbb{P}^{3N-1}$. In type IIA stringy representation, this
situation describes the case where the gauge symmetry is $U\left( 1\right)
^{N}$. For the case of $U\left( N\right) $ gauge symmetry, the D6 branes
should be coincident and so the corresponding curve classes $C_{6}^{\left(
ii\right) }$ have to be degenerate curves in $\mathbb{B}_{1}$. This
corresponds to,
\begin{equation}
NC_{6}=C_{6}^{\prime }=N\left( 3H_{1}-2E_{1}\right) ,  \label{nd6}
\end{equation}
where $H_{1}$ and $E_{1}$stand for the basic classes of the del Pezzo
surface $\mathbb{B}_{1}$ where live the degenerate $NC_{6}$.\newline
(\textbf{3}) $N$ holomorphic curves $C_{1}^{\left( 0i\right) }+C_{1}^{\left(
i0\right) }$ solved as $\left( H_{0}-E_{i}\right) +\left( H_{i}-E_{0}\right)
$ and stretching between $C_{2}^{\left( 00\right) }$ and $C_{6}^{\left(
ii\right) }$. For the case of N coincident D6 branes eq(\ref{nd6}), the N
classes $C_{1}^{\left( 0i\right) }+C_{1}^{\left( i0\right) }$ fuse and give
\begin{eqnarray}
C_{1}^{\left( 01\right) } &=&NH_{0}-E_{1}  \notag \\
C_{1}^{\left( 10\right) } &=&H_{1}-NE_{0},  \label{nd3}
\end{eqnarray}
where obviously the F1 strings stretching between $C_{2}^{\left( 00\right) }$
and the $NC_{6}^{\left( ii\right) }$ are collectively described by the sum $%
\left( NH_{0}-E_{1}\right) +\left( H_{1}-NE_{0}\right) $. From these
solutions, it not difficult to check that the constraint eqs (\ref{cst1})
are exactly fulfilled.\newline
(\textbf{4}) Finally for the $K$ D0-branes describing the quantum flux, the
construction is quite similar to what we have done for the case of
coincident D6 branes. The homology class describing the K D0 branes is $%
C_{0}^{\left( -1-1\right) }=kE_{-1}$ and the F'1 strings stretching between
the kD0 and D2 realized as $C_{0}^{\left( -1-1\right) }$ and $C_{2}^{\left(
00\right) }$\ are given by
\begin{equation}
C_{1}^{\left( -10\right) }=\left( kH_{0}-E_{-1}\right) +\left(
H_{-1}-kE_{0}\right) ,  \label{nd4}
\end{equation}
They quantum fluxes are naturally given by the ends of the F'1 strings on $%
C_{2}^{\left( 00\right) }$ and so are associated with the intersection
number $C_{1}^{\left( -10\right) }.C_{2}^{\left( 00\right) }=k$ in agreement
with the constraint eqs.

\section{Conclusion and Discussion}

Using a recent result linking p-branes and holomorphic curves in del Pezzo
surfaces, we have developed a new way to deal with brane bounds of M-theory
on $\mathbb{S}^{1}$. To illustrate our idea in an explicit manner, we have
considered the usual type IIA stringy representation of the quantum Hall
soliton (QHS) and derived its realization by using the H$_{2}$ homology of
del Pezzo surfaces. In our representation, QHS is described by a system of
intersecting classes of holomorphic curves as given by eqs(\ref{nd1}-\ref
{nd4}), see also figure 6.

The idea developed here can be used to derive new solutions for QHS but also
for studding general branes systems. The development of these issues seems
to us important, it offers an other way to approach p-brane bounds and uses
the powerful tools of homology groups and algebraic geometry that may allow
to open new horizons. In particular, one may derive new representations of
higher dimensional quantum Hall solitons involving two D4-branes and F1
strings stretching between them in the same spirit as in \cite{26,27,28}.
One may also consider QHS using p-branes of type IIB superstring that are
dual to the previous type IIA ones. In the algebraic geometry of QHS we have
been considering, this configuration can be obtained without major
difficulty. It consists of the system D3/S$^{1}$, D7/S$^{1}$, F1, D1 and D1/S%
$^{1}$ and satisfy similar constraint eqs to relations(\ref{wh}). The
correspondence between the two representations is as follows,

\begin{equation}
\begin{tabular}{|l|l|l|l|l|l|}
\hline
Type IIA & D2 & F1 & D0 & D4 & D6 \\ \hline
Curves in $\mathbb{B}_{1}$ & $H$ & $H-E$ & $E$ & $2H-E$ & $3H-2E$ \\ \hline
Type IIB & D3/S$^{1}$ & F1, D1 & F1/S$^{1}$, D1/S$^{1}$ & D5/S$^{1}$ & D7/S$%
^{1}$ \\ \hline
Curves in $\mathbb{B}_{2}$ & $l_{1}+l_{2}-e$ & $l_{1},l_{2}$ & $%
l_{1}-e,l_{2}-e$ & $2l_{1}+l_{2}-e$ & $3l_{1}+l_{2}-e$ \\ \hline
\end{tabular}
,
\end{equation}
where $l_{i}.l_{j}=1-\delta _{ij},$ $l_{i}.e=0$ and $e.e=-1$\ . To algebraic
geometry engineer the corresponding QHS dual to the type IIA one, all one
has to do is, instead of the surface $\mathbb{B}_{1}$ generated by $l_{1}$
and $l_{2}$, one considers rather the del Pezzo surface $\mathbb{B}_{2}$,
see figure 7

\begin{figure}[tbh]
\begin{center}
\epsfxsize=6cm \epsffile{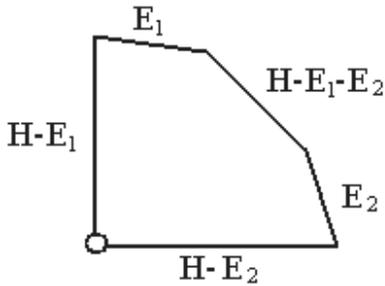}
\end{center}
\caption{This graph represents the toric diagram of the del Pezzo surface $%
\mathbb{B}_{2}$.}
\end{figure}
The extra blow up described by the exceptional class $e$ deals with the
brane wrapping cycle $\mathbb{S}^{1}$. The solution to the constraint eqs
may be obtained without difficulty by using the mapping
\begin{equation}
e=H-E_{1}-E_{2};\qquad l_{1}=H-E_{1};\qquad l_{2}=H-E_{2}
\end{equation}
Applying the rules we have used\ in elaborating the type IIA stringy
realization of quantum Hall soliton, we can draw here also the graphs of the
F1, D1 strings and the wrapped D-branes D3/S$^{1}$ and D7/S$^{1}$ involved
in the type IIB stringy representation of QHS. We have,

\begin{figure}[tbh]
\begin{center}
\epsfxsize=14cm \epsffile{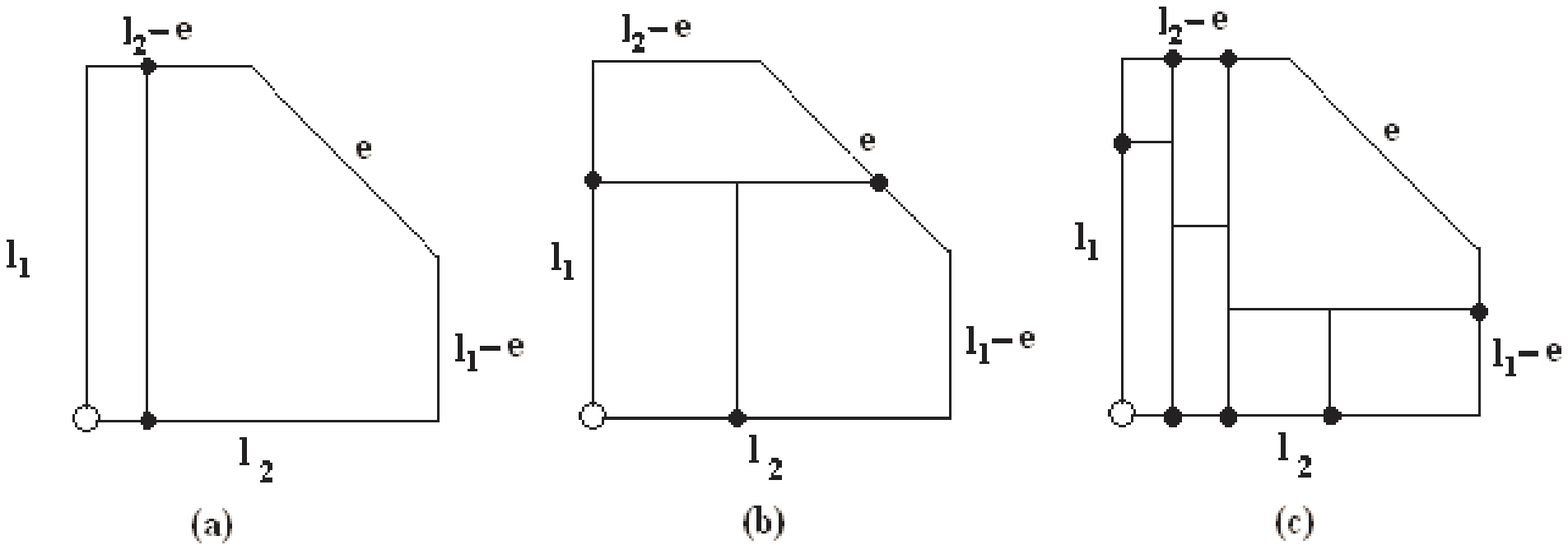}
\end{center}
\caption{\textit{These are three graphs of curves in the del Pezzo surface }$%
B_{2}$\textit{, they are involved in the type IIB stringy representation of
QHS. These classes are associated with branes in type IIB string on }$S^{1}$%
\textit{. Figure 12a gives a representation of F1 string. Figure 12b
represents a wraped D3 brane on a circle and figure 12c describes a wrapped
D7 brane on a circle.}}
\end{figure}

Using these graphs, one can also build the QHS diagram similar to that given
by figure 10. Details on this issue as well as other aspects dealing with
the derivation of new solitons including higher dimensional QHS with a
configuration type D4-F1-D4-D0 will be presented elsewhere.

\begin{acknowledgement}
Ait Ben Haddou thanks Department of Mathematics, Faculty of Meknes and M.
Zaoui for support. Abounasr, El Rhalami and Saidi thank A Belhaj, L. El
Fassi, A Jellal and E M Sahraoui for earlier collaborations on these issues.
This research work is supported by Protars III D12/25 CNRT (Rabat).
\end{acknowledgement}

\end{document}